\documentclass[final,5p,times,twocolumn]{elsarticle}

 \usepackage{graphics}

\usepackage{amssymb}

\usepackage{lineno}

\usepackage{subfigure}

\journal{Nuclear Instruments and Methods A }

\begin{document}

\begin{frontmatter}





\title{PIXEL 2010 - a R$\acute{\mathrm e}$sum$\acute{\mathrm e}$}
\author{N.~Wermes\fnref{label1}}
\ead{wermes@uni-bonn.de}
\ead[url]{http:hep1.physik.uni-bonn.de}
\fntext[label1]{Work supported by the German Ministerium f{\"u}r Bildung, Wissenschaft, Forschung und Technologie (BMBF) under contract no.~05 H09PD2.}
\address{Physikalisches Institut, Bonn University, D 53115 Bonn, Germany}

\begin{abstract}
The Pixel 2010 conference focused on semiconductor pixel detectors for particle tracking/vertexing as well as for
imaging, in particular for synchrotron light sources and XFELs. The big LHC hybrid pixel detectors have impressively
started showing their capabilities. X-ray imaging detectors, also using the hybrid pixel technology, have greatly advanced the experimental possibilities
for diffraction experiments. Monolithic or semi-monolithic devices like CMOS active pixels and DEPFET pixels have now reached a state such that complete vertex detectors for RHIC and superKEKB are being built with these technologies. Finally, new advances towards fully monolithic active pixel detectors, featuring full CMOS electronics merged with
efficient signal charge collection, exploiting standard CMOS technologies, SOI and/or 3D integration, show the path for the future.
This r$\acute{\mathrm e}$sum$\acute{\mathrm e}$ attempts to extract the main statements of the results and developments presented at this conference.
\end{abstract}


\end{frontmatter}

 \linenumbers


\section{Introduction}
The first conference on pixel detectors in particle physics was held in 1988 in Leuven. During the past 20 years pixel detectors have been developed
to precise tracking and imaging detection devices, first as hybrid pixel detectors, the technology of choice for the LHC and for new detectors
for X-ray imaging at synchrotron light sources. The advances in CMOS technology have opened possibilities towards more monolithic or semi-monolithic
devices, with trade-offs to be made between charge collection (full or uncomplete), the level of circuit integration (partial or full CMOS), and the level of 3-dimensional
integration. As of today compromises must be made. That these approaches have matured is evidenced by the fact that non-hybrid technologies are chosen for the
pixel vertex detectors of the STAR experiment at RHIC (CMOS active pixels) and of the Belle II experiment at superKEKB (DEPFET pixels).
This demonstrates the current direction of the pixel development, shown in Table~\ref{rates} which compares particle rates and fluences for vertex detector environments at the most relevant running and planned collider experiments. For the LHC and its upgrade programme, where very high radiation doses are expected,
the hybrid pixel technology still is the technology of choice to which there is no alternative. The price to pay is a comparatively high material budget of order $1.5\%$-$2\%$ x/X$_0$ per layer if not more. For accelerator at which the rates and radiation levels are lower, non-hybrid pixel technologies are chosen promising material budgets almost an order of magnitude lower than for the LHC.
\begin{center}
\begin{table*}[ht]
{\small
\hfill{}
\begin{tabular}{|c|ccccc|}
\hline
& luminosity & BX time & part. rate & fluence & ion. dose \\
& (cm$^{-2}$ s$^{-1}$) & (ns) & (kHz/mm$^2$) & (n$_{eq}$/cm$^2$) & (kGy) \\
\hline
LHC & 10$^{34}$ & 25 & 1000 & 10$^{15}$ & 790 \\
superLHC & 10$^{35}$ & 25 or 50 & 10000 & 10$^{16}$ & 5000 \\
superKEKB & 10$^{36}$ & 2 & 400 & $\sim$3 $\times$10$^{12}$ & 50 \\
ILC & 10$^{34}$ & 350 & 250 & 10$^{12}$ & 4 \\
RHIC & 8$\times$10$^{27}$ & 110 & 3.8 & 5 $\times$ 10$^{13}$ & 15 \\
\hline
\end{tabular}}
\hfill{}
\caption{Particle rates and fluences for various colliders and their experiments at the position of the innermost pixel layer. BX is the abbreviation for bunch crossing. The fluences and doses are given for the full life times, i.e. 7 years of design luminosity for LHC and sLHC (ATLAS and CMS), 5 years for RHIC (STAR) and superKEKB (Belle II), and 10 years for the ILC (ILD).}
\label{rates}
\end{table*}
\end{center}

For imaging applications, hybrid pixels have also been the prime choice so far, in particular since material considerations do not play a large role. Going from
X-ray imaging and imaging at synchrotron light sources to the new demands at the X-ray laser sources (XFEL) in Stanford and Hamburg, the challenges are
identified by (a) a huge dynamic range of photon flux (up to 10$^6$), (b) high count rates per pixels ($>$ MHz), (c) very large frame rates ($\sim$ 5 MHz), and (d)
little dead time and a 'seamless' architeture.

\section{The LHC Pixel Detectors}
\begin{figure}[h!!!]
\begin{center}
\subfigure[dE/dx measurement (ATLAS)]{
\includegraphics[width=0.4\textwidth]{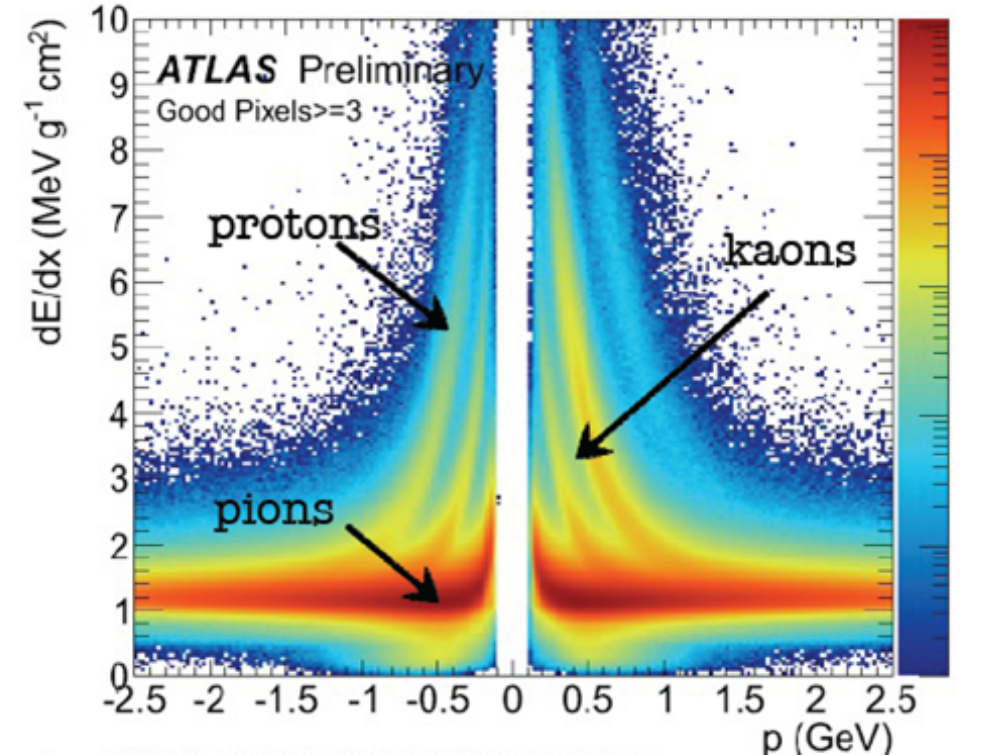}
    \label{fig:dEdx}}
    \vskip 0.5cm
\subfigure[space resolution: CMS]{
    \includegraphics[width=0.4\textwidth]{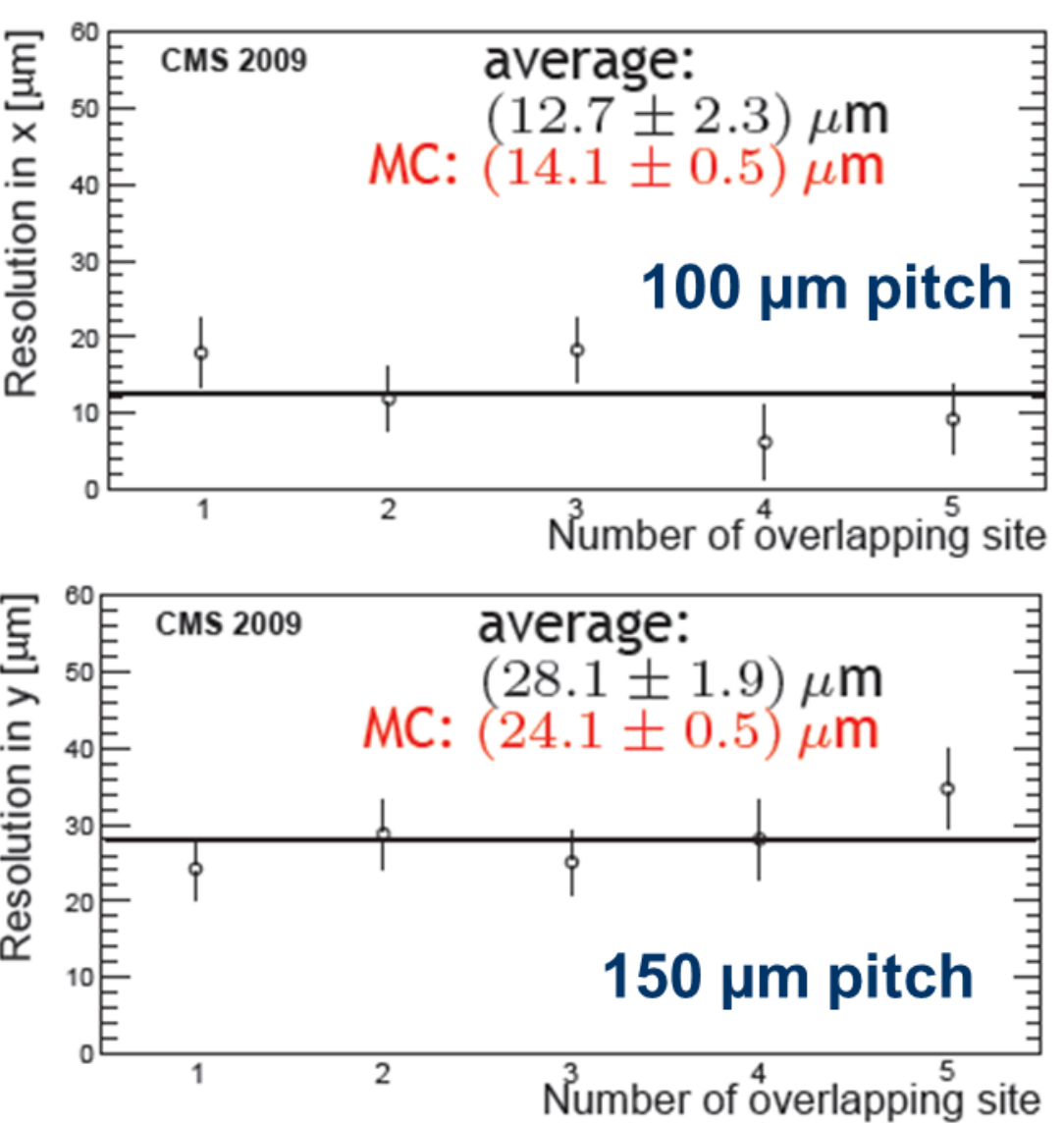}
    \label{fig:CMS-resol}}
\end{center}
\caption[]{\label{Q-measurement} (a) Measurement of the specific energy loss (dE/dx) with the ATLAS pixel detector~\cite{dellasta_pixel2010}, (b) CMS spatial resolution in the 100 $\mu$m pitch direction (top) and in the 150 $\mu$m pitch direction (bottom) as a function of the number of overlapping detector sites along the trajectory of a track~\cite{langenegger_pixel2010}.}
\end{figure}

To underline it: these are {\it big} pixel detectors: 2-3 layers/disks, roughly one meter long ($\sim$7m with services), of order 10$^8$ individually amplified channels.
In my view the main message presented at this conference on the operation experience in LHC collisions with these big pixel detectors  is~\cite{moss_pixel2010,bolla_pixel2010,keil_pixel2010,kreis_pixel2010,langenegger_pixel2010,riedler_pixel2010,dellasta_pixel2010}: They are essentially {\it noiseless} devices. With a noise hit level per pixel and bunch crossing below $10^{-9}$ after masking pathological channels and after reconstruction, less than 0.2 noise hits   per event are observed~\cite{dellasta_pixel2010}. Operated at modest threshold settings ($\sim$3000 e$^-$ - 4000 e$^-$), efficiencies above 99$\%$ and trigger rate capabilities up to 80 kHz have been
achieved. This must be considered a great achievement which resulted from many years of pixel R$\&$D. The resolution of the charge measurement is excellent as is demonstrated for low momentum tracks ($<$ 1GeV) with large specific energy loss in Fig.~\ref{fig:dEdx}. The only drawback that one can identify is the not so small
amount of material that these trackers have, more than 3$\%$ of a radiation length per layer in the present trackers, still well above 1$\%$ for current sLHC
designs. This is largely due to the fierce radiation environment at LHC requiring
the hybrid pixel technology and the detectors to be operated at temperatures below 0$^\circ$, added by the fact that they are a ´'first of their kind`'. The ALICE pixel detector is an exception achieving a 1.1$\%$ x/X$_0$ per layer due to a dedicated effort in minimizing the material in all aspects. This is helped by the fact that for heavy ion collisions at LHC a homogeneous 4$\pi$ coverage is not mandatory as it is in pp collisions. Therefore the stiffness of the very thin support structure can be obtained by radial, chamber-type structure elements. Furthermore, the lower luminosity for heavy ion collisions results in less detector radiation damage which in turn allows to operate the pixel detector at room temperature saving cooling material.

An interesting aspect which is now seen with the first data is the effect of the different charge sharing tunings of ATLAS and CMS in comparison. ATLAS with
50 $\times$ 400 $\mu$m$^2$ pixel cell dimensions has chosen a module tilt angle such that it largely compensates the Lorentz angle inclination of the charge drift.
This maximizes the seed pixel charge, an important aspect after irradiation induced charge loss, but not optimal for space resolution.
CMS instead has tuned for optimal charge sharing of its 100 $\times$ 150 $\mu$m$^2$ pixels in the 100 $\mu$m direction. This leads to an excellent resolution of
$\sigma_x = (12.7 \pm $2.3) $\mu$m~\cite{langenegger_pixel2010} for 100 $\mu$m pixel pitch (with a binary resolution of roughly 30 $\mu$m) as shown in Fig.~\ref{fig:CMS-resol} (top), almost as good as ATLAS ($\sigma_x \lesssim$ 10 $\mu$m, $\sigma_z \approx$ 115 $\mu$m) obtained with 50 $\mu$m pitch. The charge sharing is less effective in the orthogonal z-coordinate (150 $\mu m$ pitch) where $\sigma_z = (28.1 \pm $1.9) $\mu$m are obtained compared to 45 $\mu$m binary resolution. Increasing irradiation will, as a consequence, deteriorate the resolution more strongly for CMS~\cite{rohe_pixel2010}.

The data taken so far with the three large LHC pixel detectors clearly justify the huge R\&D effort that went into their development. The clean measurements of the beam spot position and its tilt, a complete mapping of materials by conversion finding and clean measurements of particle resonances like the $J/\psi$, important for the detector calibration, have been shown at this conference~\cite{hirsch_pixel2010,dinardo_pixel2010}.

\begin{figure}[thb]
\begin{center}
\subfigure[planar Si sensors (n in p)]{
\includegraphics[width=0.4\textwidth]{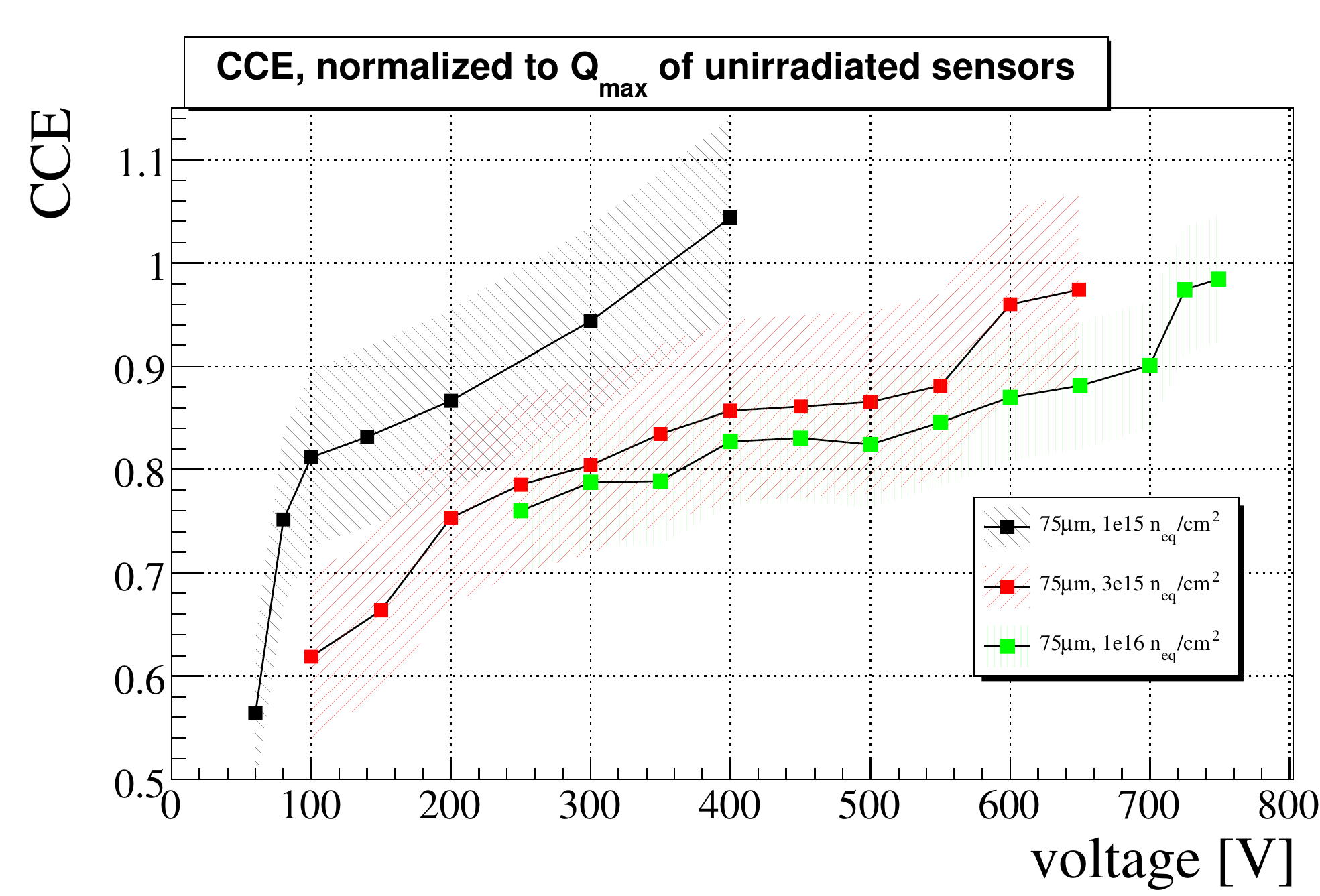}
    \label{fig:planar}}
 \vskip 0.5cm
\subfigure[Thinned Chip and Through Silicon Via]{
    \includegraphics[width=0.4\textwidth]{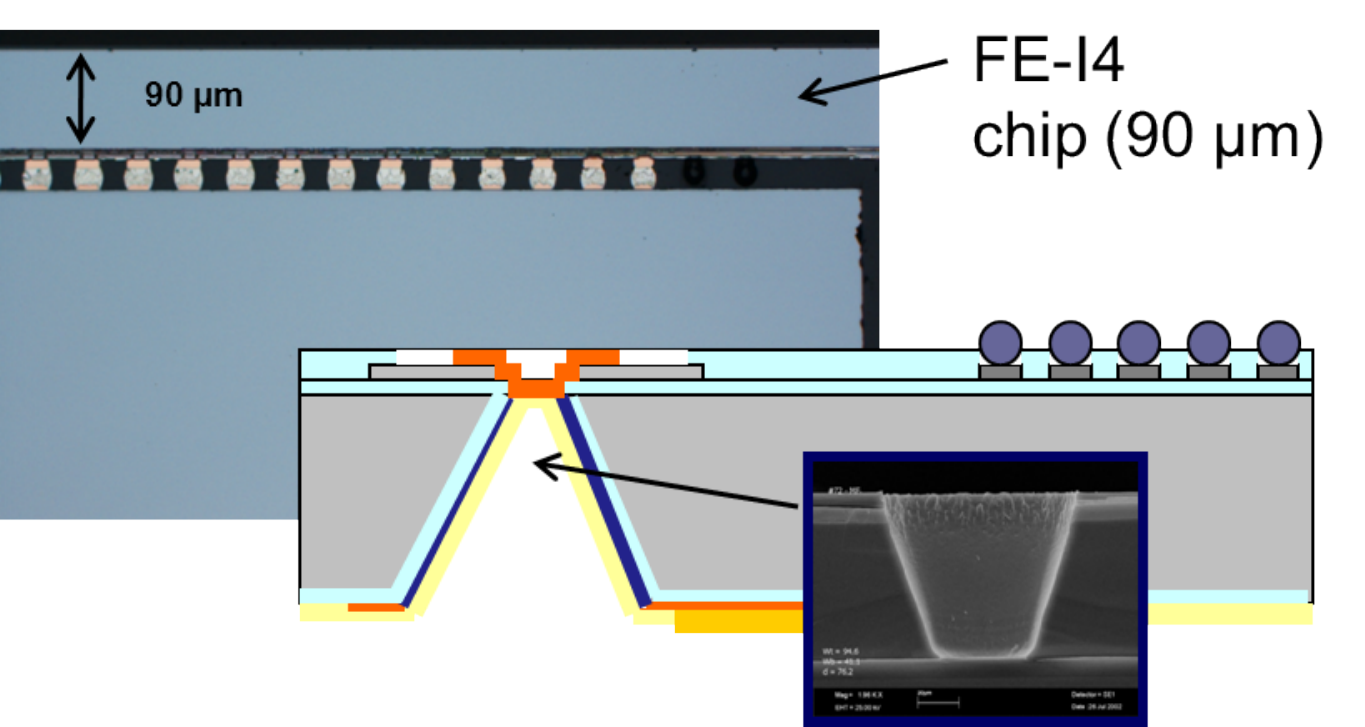}
    \label{fig:TSV}}
\end{center}
\caption[]{\label{Q-collection} (a) Charge collection efficiency (CCE) of n in p planar silicon pixel sensors (75 $\mu$m thick) as a function of bias voltage after irradiation. Low cost flip chip assembly and interconnection technologies for to 1 (black), 3 (red) and 10 (green) $\times$ 10$^{15}$ n$_{eq}$ cm$^{-2}$~\cite{macchiolo_pixel2010}. (b) SEM photograph of a bumped chip-sensor assembly with a thinned (90 $\mu$m) large area FE chip and sketch of a tapered TSV with SEM photo~\cite{gonella_pixel2010}.}
\end{figure}

The LHC pixel collaborations have started to plan for upgrades. While CMS plans for a completely new pixel detector with 4 barrel layers and 2 $\times$ 3 disks~\cite{bean_pixel2010}, ATLAS will add an innermost layer to be inserted into the existing 3-layer detector exploiting the usage of a smaller beam pipe of 3 cm radius~\cite{hugging_pixel2010}. Later ($\sim$2020) completely new tracking detectors are planned by both experiments.
A major goal is to half the material, i.e. 1.5 $\%$ x/X$_0$ average per layer. To achieve this, apart from using light weight
new material support structures, large efforts go into thinning of bumped chips~\cite{gonella_pixel2010} and new routing schemes exploiting 3D techniques and through silicon vias (TSV)~\cite{gonella_pixel2010, fritzsch_pixel2010, macchiolo_pixel2010} which allow routing of R/O lines also on the backside of chips. Figure~\ref{fig:TSV} shows a successful thin chip assembly and a tapered TSV structure which is easier to realize on
thin chips~\cite{gonella_pixel2010, fritzsch_pixel2010}.
To address the expected increased rate by a factor more than 3 and up to 10 compared to the design luminosity, a new front-end chip (FE-I4)
has been designed~\cite{barbero_pixel2010} that can cope with the increased rate and needed bandwidth with excellent efficiency. For the sensors three options (planar Si, 3D silicon, CVD diamond) are studied which all have their advantages and disadvantages. Planar silicon sensors (n in n as well as n in p)~\cite{munstermann_pixel2010} are best understood and are  much lower in cost compared to other options. They require, however, voltages in excess of $\sim$1000 V for full depletion after irradiation to 10$^{16}$ n$_{eq}$ cm$^{-2}$ and a new slim (to maximize the active area) guard ring design bringing down the voltage at the sensor edges. A new study~\cite{macchiolo_pixel2010} has shown that thin n in p planar sensors still have good charge collection efficiency (CCE) at voltages well below 1000 V (see Fig.~\ref{fig:planar}). 3D silicon sensors with vertical electrodes etched into the bulk~\cite{micelli_pixel2010}, in contrast, promise high radiation tolerance at low voltages and large active area at the expense of a much less developed sensor production process and some loss in charge collection efficiency in the vertical highly doped columns as well as larger noise than planar sensors. The main issue currently is the production yield. Diamond pixel detectors have a $\sim$3 times smaller signal at the same thickness due to the larger band gap than Si, but they offer zero leakage currents  even after strong irradiation and smaller pixel capacitances and hence lower noise, such that the S/N ratio of diamond after irradiation is, in fact, very competitive to silicon.
Very attractive is the simultaneous excellent heat conduction capability of diamond which could be exploited for future trackers. As of today the question remains if one can produce diamond detectors, especially single crystal CVD, in sufficient quantity on a reasonable time scale.
The tremendous success of the hybrid pixel technology in particle tracking has found followers in new experiments like PANDA at the FAIR facility in Germany~\cite{stockmanns_pixel2010} and to allow tracking in low background experiments like COBRA~\cite{schwenke_pixel2010} to further enhance the background suppression in 0$\nu\beta\beta$ decay search as was reported at this conference.

\section{New Pixel Trackers for New Colliders}
\begin{figure}[thb]
\begin{center}
\subfigure[DEPFET module in frame]{
\includegraphics[width=0.43\textwidth]{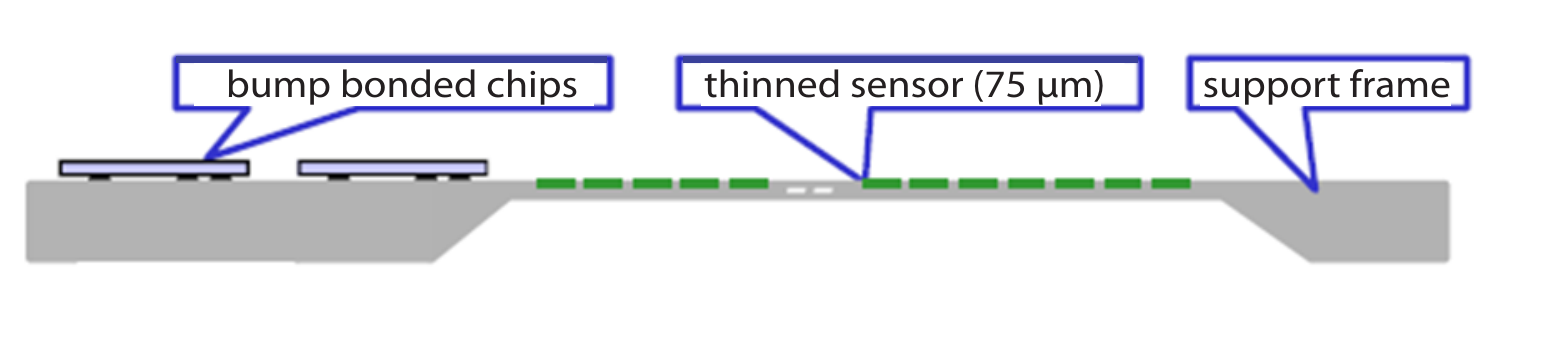}
    \label{fig:DEPFET-frame}}
  \hskip -0.3cm
  \vskip 0.5cm
\subfigure[DEPFET prototype module with steering and R/O chips on PCB board]{
    \includegraphics[width=0.40\textwidth]{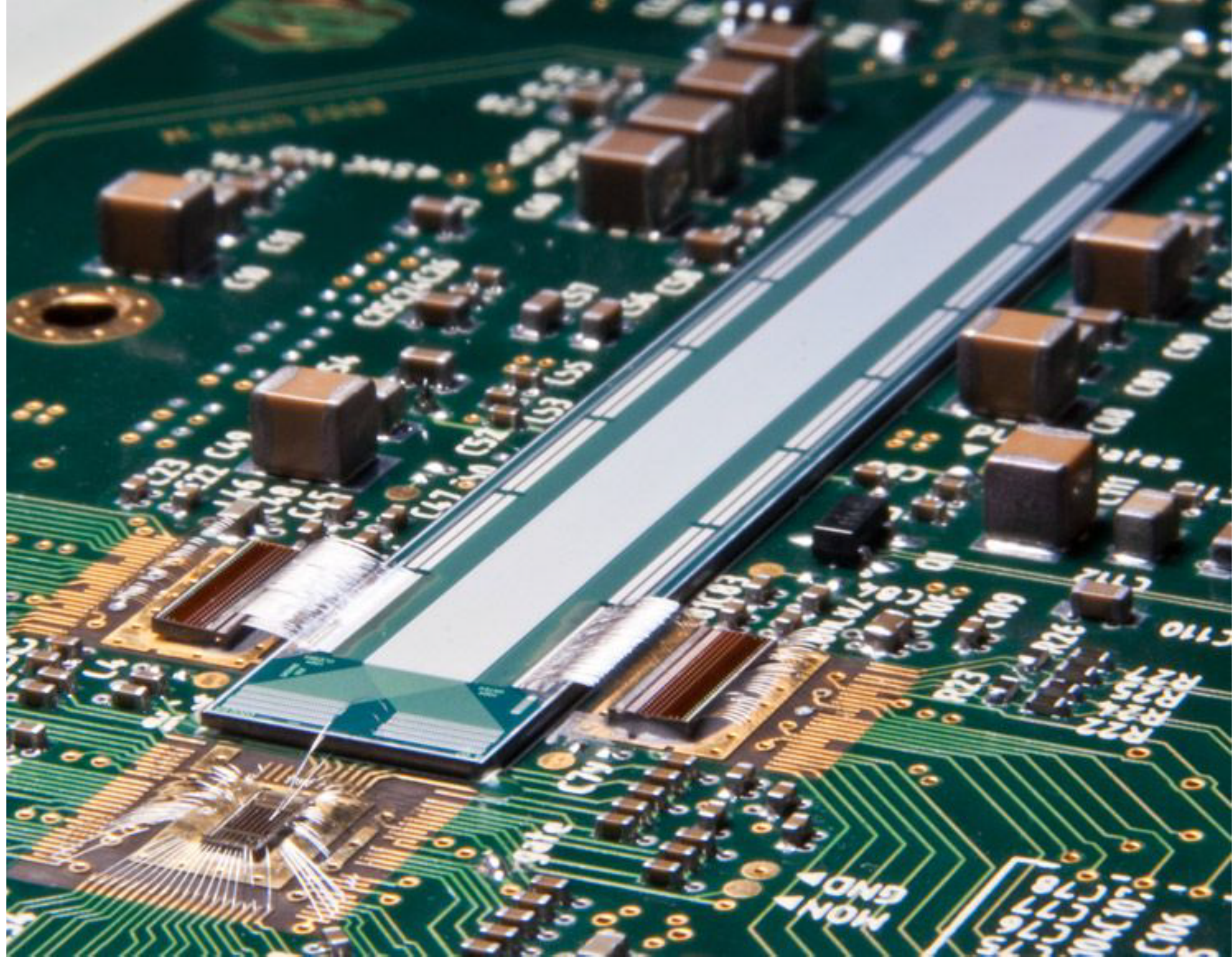}
    \label{fig:DEPFET-module}}
\end{center}
\caption[]{\label{fig:DEPFET} (a) DEPFET module with thinned sensor in support frame. (b) Photograph of a DEPFET prototype module~\cite{marinas_pixel2010}.}
\end{figure}

At rates somewhat lower and fluences as well as radiation doses much lower than at LHC and sLHC, active CMOS pixels and DEPFET pixels are very suitable devices for vertex detectors and are in fact planned at the heavy ion accelerator RHIC for the STAR-experiment \cite{greiner_pixel2010,dokhorov_pixel2010} and at superKEKB for Belle II~\cite{marinas_pixel2010} or at SuperB~\cite{rizzo_pixel2010}. Current generation CMOS pixel detectors like MAPS~\cite{meynants98_1,MAPS-Turchetta} allow electronics circuitry (mostly nMOS) within the active sensor area on the expense of diffusion based (small) signal charge collection, DEPFET pixels~\cite{kemmerandlutz88} have an amplification transistor implanted in a fully depleted bulk. Both these approaches differ much from the above mentioned hybrid pixel technology employed at LHC. In particular, their readout is frame based with sequential row selection and parallel column readout provided by readout chip electronics located at the sensor edges rather than flipped atop the sensors. This results in the following common advantages: large area bump bonding and IC material in the active area is avoided, the active area of the sensor can be made very thin ($\sim$50 $\mu$m) resulting in a total material budget in the order of $0.2 \%\ x/X_0$. The consumed power is low due to the fact that only one or two rows is active at a time and hence less cooling is needed. Very small pixel linear dimensions (${\cal{O}}$(20$\mu$m)) are possible. For Belle II (DEPFET) the pixel size is limited by the data bandwidth which enormously increases with very small and hence very many pixels for the same area, such that larger pixels (${\cal{O}}$(50$\mu$m)) are chosen. On the down side, MAPS and DEPFET technologies suffer from the larger vulnerability to radiation and lower readout speed compared to the hybrid pixel technology (see Table~\ref{rates}). Figure~\ref{fig:DEPFET} shows as an example of these non-hybrid detectors the DEPFET module concept for Belle II which employs backside etching for thickness reduction with the mechanical strength being supplied by the remaining frame~\cite{marinas_pixel2010}. Also shown at this conference was a new approach for efficient cooling in a SuperB pixel vertex detector using micro channels~\cite{bosi_pixel2010}, noting that the heat transfer coefficient is inversely proportional to the hydraulic diameter of a cooling channel.
Finally, planning and R$\&$D for a future Linear Collider~\cite{battaglia_pixel2010} targets material thicknesses and spatial resolutions
even beyond what was discussed here so far, in order to achieve impact parameter resolutions of below $\sim$10 $\mu$m even at low ($\sim$ 2 GeV) momenta.
The dedicated PLUME collaboration~\cite{Nomerotski_pixel2010} investigates ultra low weight support structures in this context.

\section{Imaging with Pixel Detectors at Synchroton Light Sources}
\begin{figure}[hbt!]
\begin{center}
\subfigure[Illustration of a diffraction experiment]{
\includegraphics[width=0.4\textwidth]{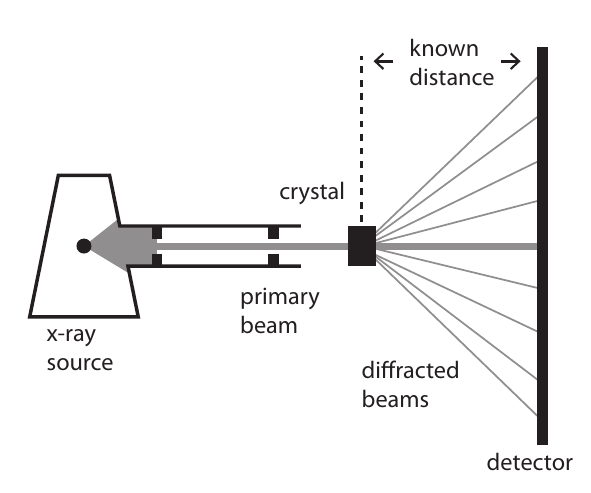}
    \label{fig:diffraction}}
\vskip 0.5cm
\subfigure[From light sources to XFELs]{
    \includegraphics[width=0.40\textwidth]{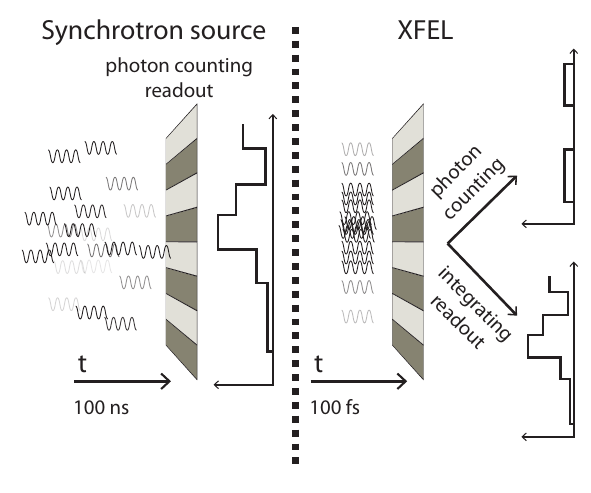}
    \label{fig:SLS-XFEL}}
\end{center}
\caption[]{\label{fig:syli} (a) Diffraction is a typical example of experiments at synchrotron light sources (illustration). (b) Sketch of the different beam situations at a synchrotron light source compared to an X-ray free electron laser situation~\cite{mozzanica_pixel2010}.}
\end{figure}

\begin{figure}[hbt!]
\begin{center}
\includegraphics[width=0.45\textwidth]{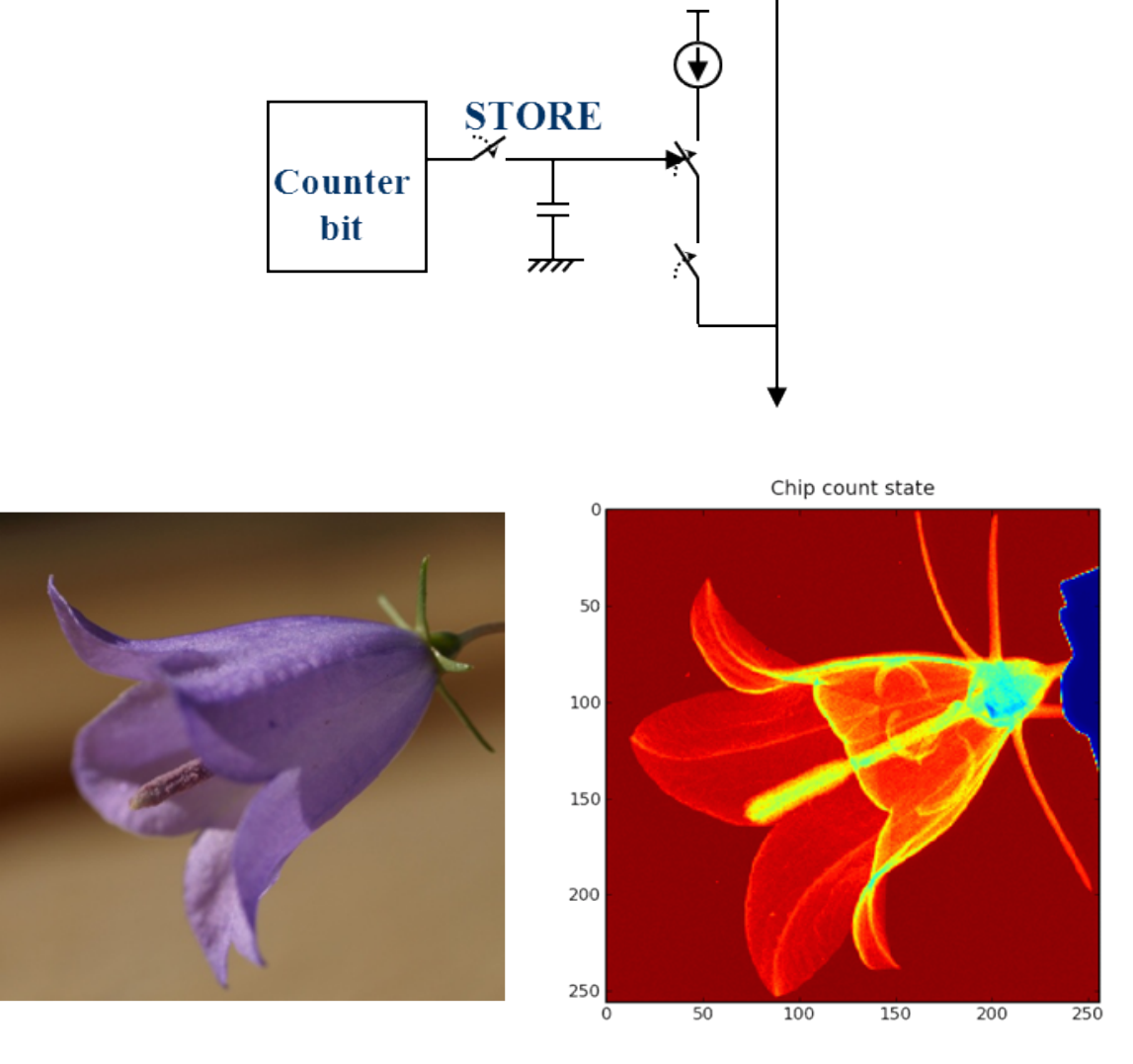}
\end{center}
\caption[]{\label{fig:EIGER} (top) Double buffering cell within the EIGER-chip~\cite{dinapoli_pixel2010} allows continuous, quasi dead time free readout in synchrotron light imaging with hybrid pixel detectors. (bottom) Real and X-ray image of flower obtained with EIGER~\cite{dinapoli_pixel2010}.}
\end{figure}

Huge progress was reported at this conference regarding pixel detector imaging in synchrotron light experiments. One application is sketched in
Fig.~\ref{fig:diffraction} -- a typical diffraction experiment.
The development of hybrid pixel detectors has also largely enriched the possibilities in synchrotron light imaging. To stay with the example, the detection of diffraction patterns requires complete area imaging of spots of different intensity with high picture rates. This in turn for counting pixel detectors means: $>$ MHz rates per pixel, frame rates above kHz at a light source, 5 MHz at an XFEL, with no or little dead time, very large dynamic range (up to 10$^6$), small pixel size ($<$ 50$\mu$m), and {\it seamless} detectors with no dead areas. The PILATUS detector~\cite{PILATUS, schulze-briese_pixel2010} has pioneered the use of pixel detectors for synchrotron light imaging with great success as reported at this conference~\cite{schulze-briese_pixel2010}. With this device, which is now also commercially available, the dynamic range for photon detection has been increased from 15-16 bit (CCD detectors) to 20 bit. Continuous shutter-free data collection with the possibility of angle-slicing and simultaneous high and low exposure data taking at room temperature have been made possible with PILATUS. At count rates of several MHz per pixel, the frame rates are
still fairly low (12.5 Hz) with 3 ms dead time.
The next generation of a hybrid pixel synchrotron light imager is EIGER~\cite{dinapoli_pixel2010}. The functioning readout chip prototype has been presented~\cite{dinapoli_pixel2010}, featuring smaller pixels (75 $\times$ 75 $\mu$m$^2$) and thus larger count rate per mm$^2$, much larger frame rates ($>$ 20 kHz),
and continuous, quasi dead time free readout. A key feature is the buffering of a complete frame (see Fig.~\ref{fig:EIGER}) while continued readout takes place.
Figure~\ref{fig:EIGER} also shows an X-ray image of a flower taken with EIGER.
Energy-selective synchrotron light imaging has been reported from Spring-8 experiments in Japan using a threshold scanning method of PILATUS~\cite{toyokawa_pixel2010}
or by a new pixel device using CdTe as sensor material and a readout chip with a window comparator selecting an energy window~\cite{hirono_pixel2010}.

\begin{figure}[h!!!bt]
\begin{center}
\subfigure[Concept of AGIPD]{
\includegraphics[width=0.4\textwidth]{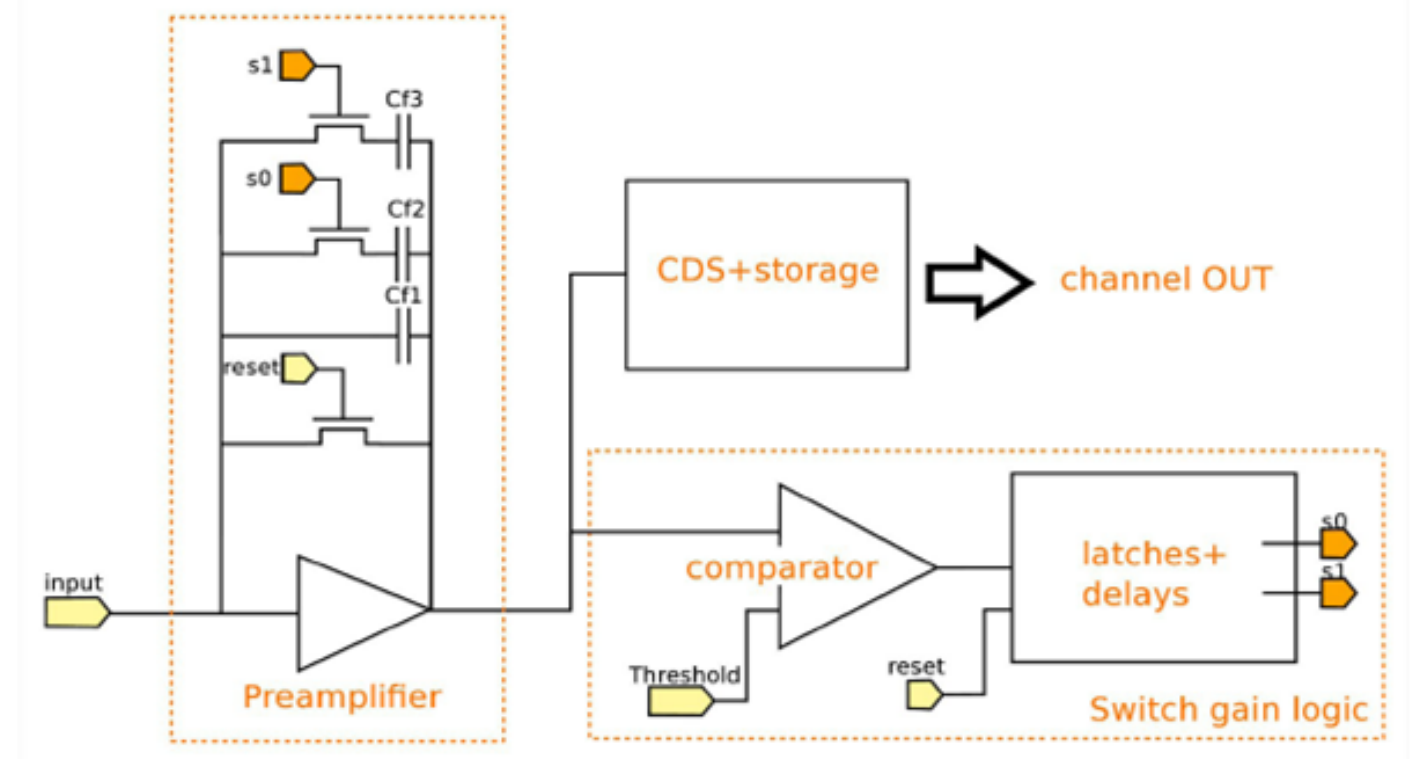}\label{fig:AGIPD1}}
\subfigure[AGIPD output characteristic (illustration)]{
\includegraphics[width=0.35\textwidth]{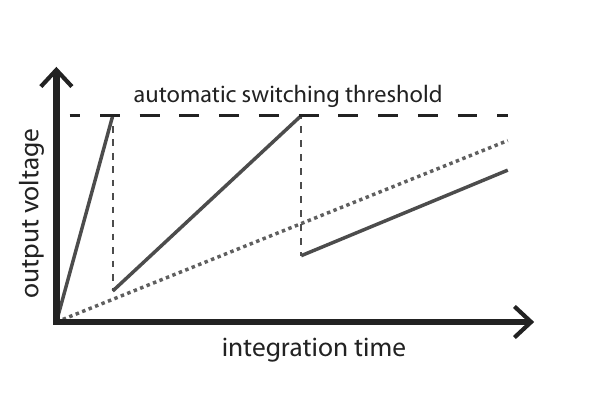}\label{fig:AGIPD2}}
    \vskip 0.3cm
\subfigure[Concept of DSSC]{
    \includegraphics[width=0.25\textwidth]{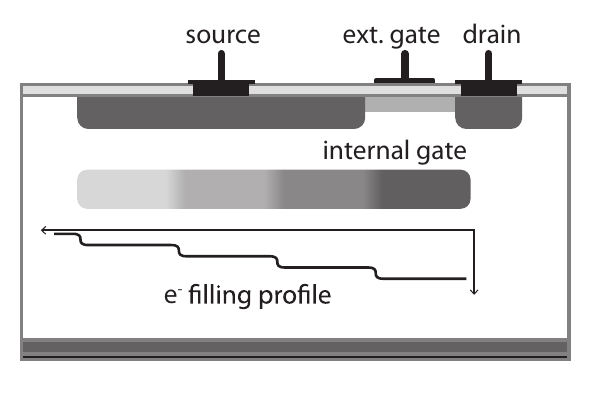}
    \includegraphics[width=0.25\textwidth]{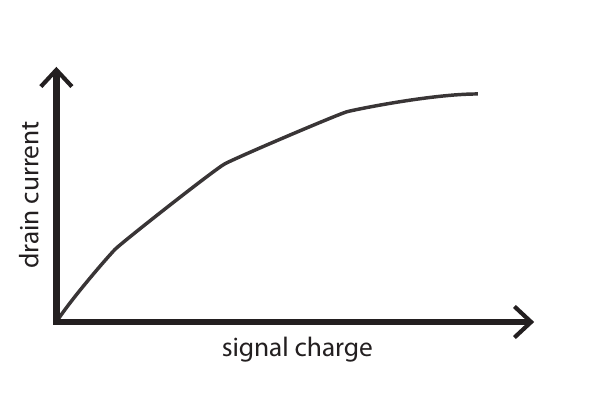}
    \label{fig:DSSC}}
\end{center}
\caption[]{\label{fig:XFELs} Two concepts to cope with a dynamic range in excess of 10$^5$ in photon detection at XFELs: (a) AGIPD concept using switched gains in the preamplifier~\cite{mozzanica_pixel2010} to deal with (b) high flux photon spots (solid) and low flux spots (dashed). (c) DSSC: signal compression by non-linear transistor characteristics using a DEPFET transistor~\cite{porro_pixel2010}, (left) sketch of the coupling of the internal gate to the DEPFET transistor, (right) resulting non-linear transistor characteristic, output current versus input charge.}
\end{figure}

\begin{figure}[ht!]
\begin{center}
\includegraphics[width=0.3\textwidth]{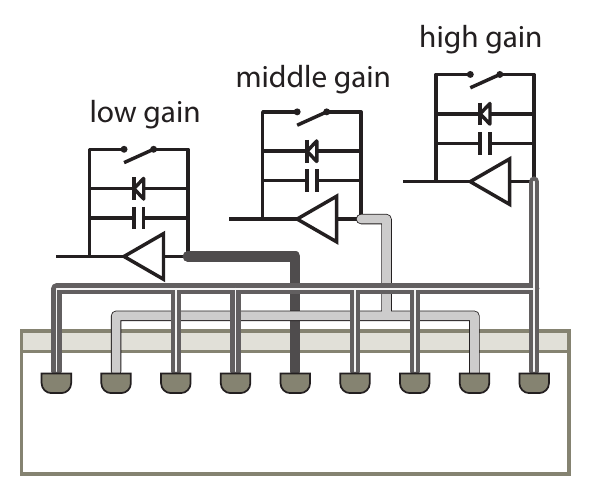}
\end{center}
\caption[]{\label{fig:XFEL-SOI} Possibilities using the SOI technology~\cite{hatsui_pixel2010} to address the large dynamic range in photon flux at XFELs. Both, different numbers of charge collection nodes and different gain stages are employed (simplified 1D illustration of a 2D array).}
\end{figure}

What is different regarding photon detection when doing experiments at an X-ray free electron laser (XFEL)? This is illustrated in Fig.~\ref{fig:SLS-XFEL}. While in conventional synchrotron light sources the X-ray photons arrive in a large number of photon bunches with low intensity irregularly distributed in time, in an XFEL they arrive coherently within extremely short 100 fs bunches separated in time by 200 ns within bunch trains. This leads to the challenge to cope with a huge dynamic range detecting from few up to 10 000 photons per 200 ns which translates into frame rates of 5 MHz. Imaging at XFELs thus requires new ways in either counting or integration
detector readout concepts. Current approaches use a switching characteristic for counting devices (GOTTHARD, AGIPD)~\cite{mozzanica_pixel2010} or a very non-linear
device characteristic leading to signal compression in the integrating DSSC-device~\cite{porro_pixel2010}. Figure~\ref{fig:XFELs} illustrates both concepts. While for the
counting readout of GOTTHARD (strips) or AGIPD (pixels) the amplifier gain is switched in a threshold driven way leading to different slopes of the characteristics when the photon flux is high (solid lines in Fig.~\ref{fig:AGIPD2}) or low (dashed lines), for the integrating DSSC device the DEPFET pixel concept mentioned earlier is employed~\cite{porro_pixel2010}. In the DEPFET device, the internal electron collecting gate steers the channel current of the implanted transistor, the gain depending on the (capacitive) coupling of the internal gate to the channel. Hence, in a geometrically large and appropriately shaped internal gate obtained by a decreasing n-doping profile (see Fig.~\ref{fig:DSSC}(left)) the current gain depends non-linearly on the filling level leading to the characteristic sketched in Fig.~\ref{fig:DSSC}, right.
Another interesting R$\&$D towards the required large photon dynamic range at XFELs was presented~\cite{hatsui_pixel2010} which exploits the possibilities
of the SOI technology. As sketched in Fig.~\ref{fig:XFEL-SOI} both, different combination of the charge collecting nodes and different gain stages (low, middle, high) are used.

Detector development in photon science at the various X-ray sources (FLASH, SCSS, LCLS, FERMI, XFEL) is being addressed within the newly formed CFEL
organisation~\cite{porro2_pixel2010} in Hamburg, Germany and first imaging experiments at LCLS using modified pnCCDs of MPI-Munich~\cite{strueder2010} were
reported~\cite{porro2_pixel2010}.

\section{News Towards Fully Monolithic Pixel Devices}
A fully monolithic pixel device, featuring full charge collection by drift in a depleted bulk, full CMOS electronics in the active area and perhaps even
3D integration allowing stacking to additional layers of electronics functionality, has been the goal of intense R$\&$D in the past decade with remarkable successes.
The Monolithic Active Pixel Sensors (MAPS) concept~\cite{MAPS-Turchetta} has paved the way so far, but this technology also suffers from small and incomplete signal charge collected by diffusion rather than drift and by the fact that the charge collecting n-well must compete with other n-wells if pMOS transistors are included in the active area, i.e. full CMOS electronics has not yet been achieved over the entire active area. To address this the {deep n-well}~\cite{deepnwell},\cite{rizzo_pixel2010} process pursued by the Pisa group creates a large n-well which acts as the charge collecting diode and has an imbedded p-well housing the nMOS transistors while the pMOS transistors are in an n-well which is geometrically smaller and less deep than the charge collecting deep n-well. Further progress on this approach using the small 65 nm technology has been reported at this conference~\cite{gaioni_pixel2010}. The 180 nm INMAPS quadruple well process used by the Rutherford group~\cite{stanitzki_pixel2010} employs a deep p-well placed underneath the pMOSTs containing n-well thus shielding it from acting as a charge drain. While these improvements yield already larger S/N ratios than with standard MAPS, the diffusion based charge collection and the only moderate radiation resistance has provoked further studies. Reported here~\cite{dokhorov_pixel2010, stanitzki_pixel2010} were attempts with a larger charge collecting diode and a new process feature providing a higher resistivity ($\sim$1 k$\Omega$cm) epitaxial layer for faster and more efficient charge collection. Charge collection efficiencies close to 100$\%$, S/N values of $\sim$30~\cite{dokhorov_pixel2010} and $\sim$90~\cite{stanitzki_pixel2010}, respectively, have been obtained. In addition, the radiation tolerance was improved by a factor of about 100~\cite{stanitzki_pixel2010}.

\begin{figure}[hbt]
\begin{center}
\subfigure[Monolithic active pixel detector in HV technology]{
\includegraphics[width=0.4\textwidth]{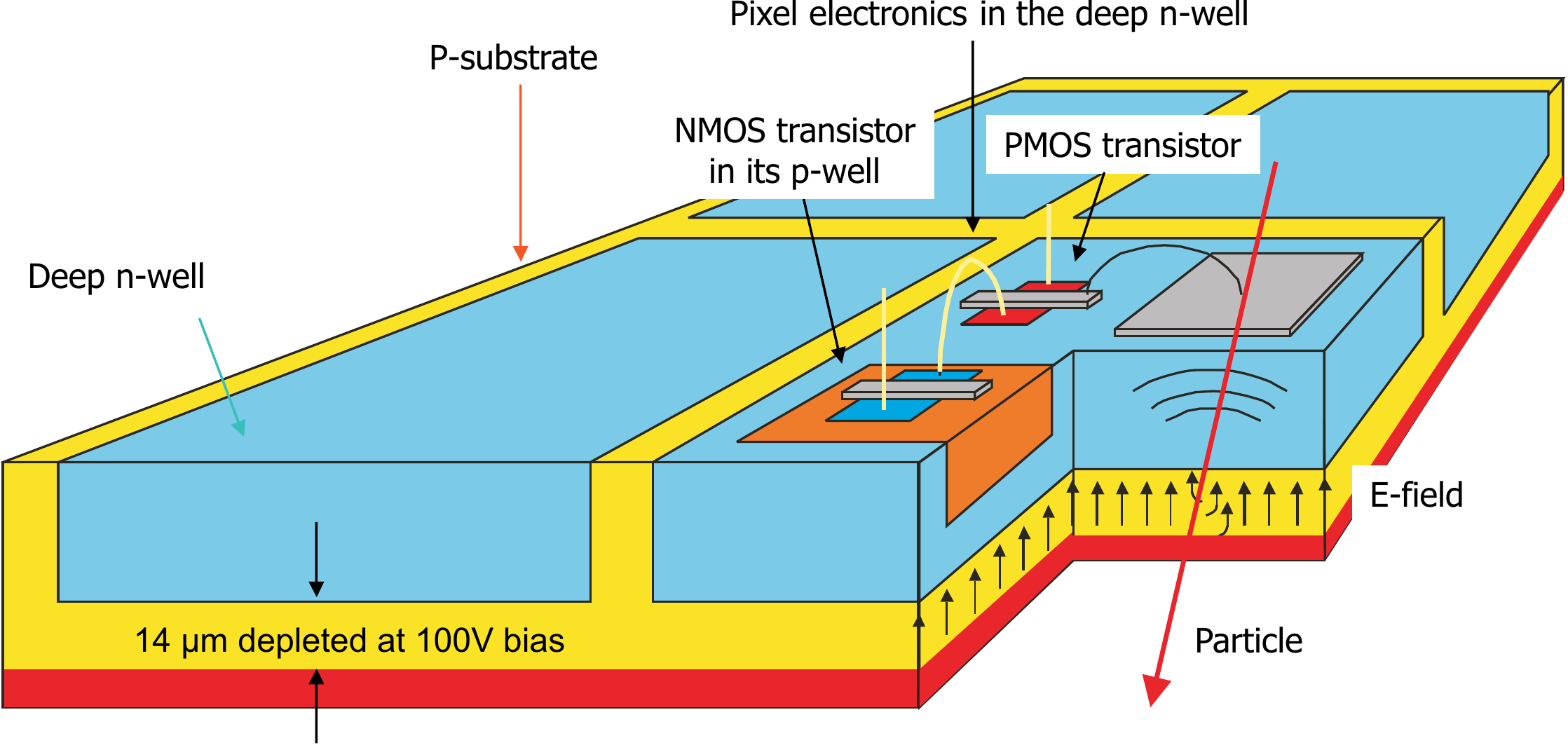}
    \label{fig:peric1}}
    \vskip 0.3cm
\subfigure[$^{55}$Fe spectrum before (left) and after (right) 10$^{15}$n$_{eq}$cm$^{-2}$]{
    \includegraphics[width=0.48\textwidth]{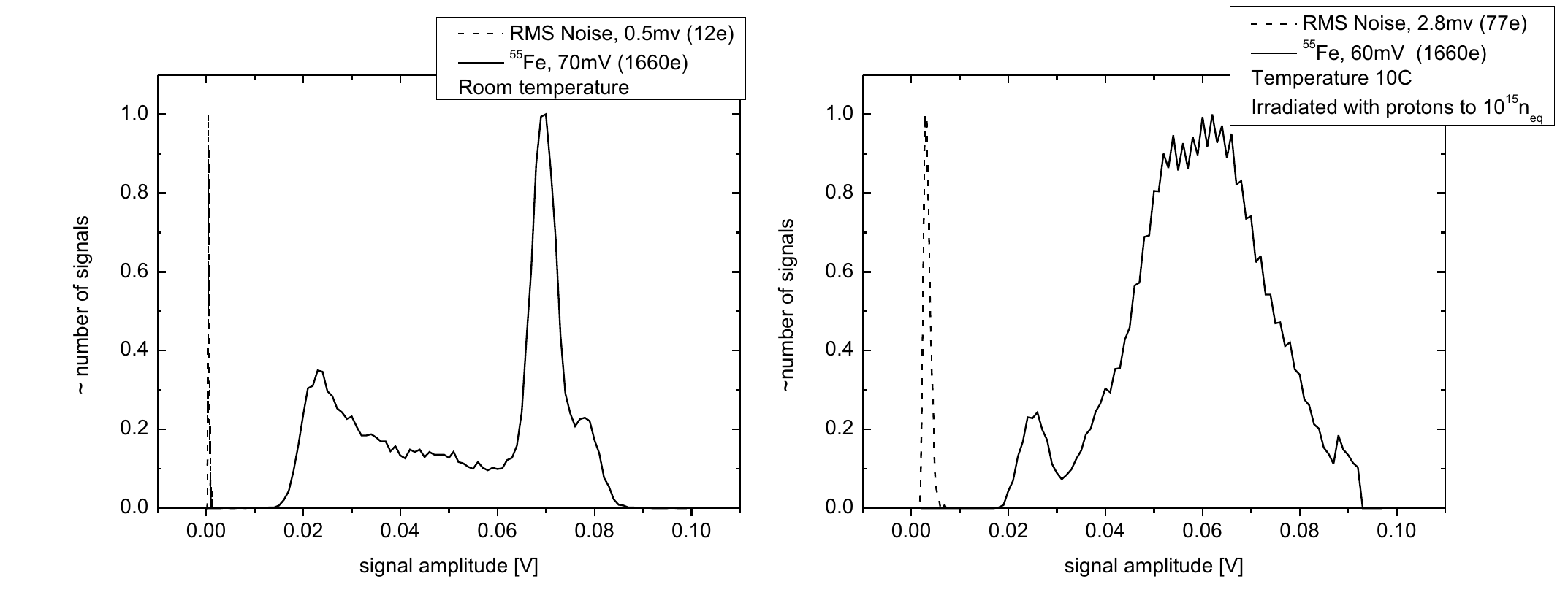}
    \label{fig:peric2}}
\end{center}
\caption[]{\label{fig:peric} Monolithic pixel detector designed in the 0.35 AMS HV process~\cite{peric2007, peric_pixel2010}. (a) Charge collection is obtained by drift in a depleted bulk. (b) Response to a $^{55}$Fe-source (6 keV $\gamma$) before (left) and after (right) irradiation to a fluence of 10$^{15}$n$_{eq}$cm$^{-2}$.}
\end{figure}

In a completely new approach~\cite{peric2007}, fairly complete characterization results have been reported~\cite{peric_pixel2010} at this conference which are quite striking. Figure~\ref{fig:peric1} shows the principle. The AMS 0.35 $\mu$m HV technology employing high-voltage n-wells in a p-substrate is used to create a monolithic pixel sensor which features full charge collection by drift in a directional E-field, 100$\%$ fill factor without charge loss due to embedding the entire structure (nMOST and pMOST in n-wells) in a deep n-well which also is the collecting diode. The device is radiation hard up to fluences of 10$^{15} $n$_{eq}$~cm$^{-2}$ as shown in Fig.~\ref{fig:peric2} which shows the response to a $^{55}$Fe source (6 keV X-ray) before and after high fluence irradiation (T=10$^o$C). The signal is widened but the irradiated spectrum (Fig.~\ref{fig:peric2}~(right)) shows a comfortable distance between the narrow pedestal peak and the signal of 1660 e$^-$.
This way many of the goals mentioned above for a monolithic detector are met: fast and $\sim$full charge collection, true CMOS circuity inside the active area, large S/N ratio ($\sim$~100), small pixel size (21 $\times$ 21 $\mu$m$^2$), and high radiation hardness (to 10$^{15}$n$_{eq}$~cm$^{-2}$), at a power of about 12 $\mu$W per pixel~\cite{peric_pixel2010}.

\begin{figure}[hbt]
\begin{center}
\subfigure[Principle of an SOI pixel detector]{
\includegraphics[width=0.4\textwidth]{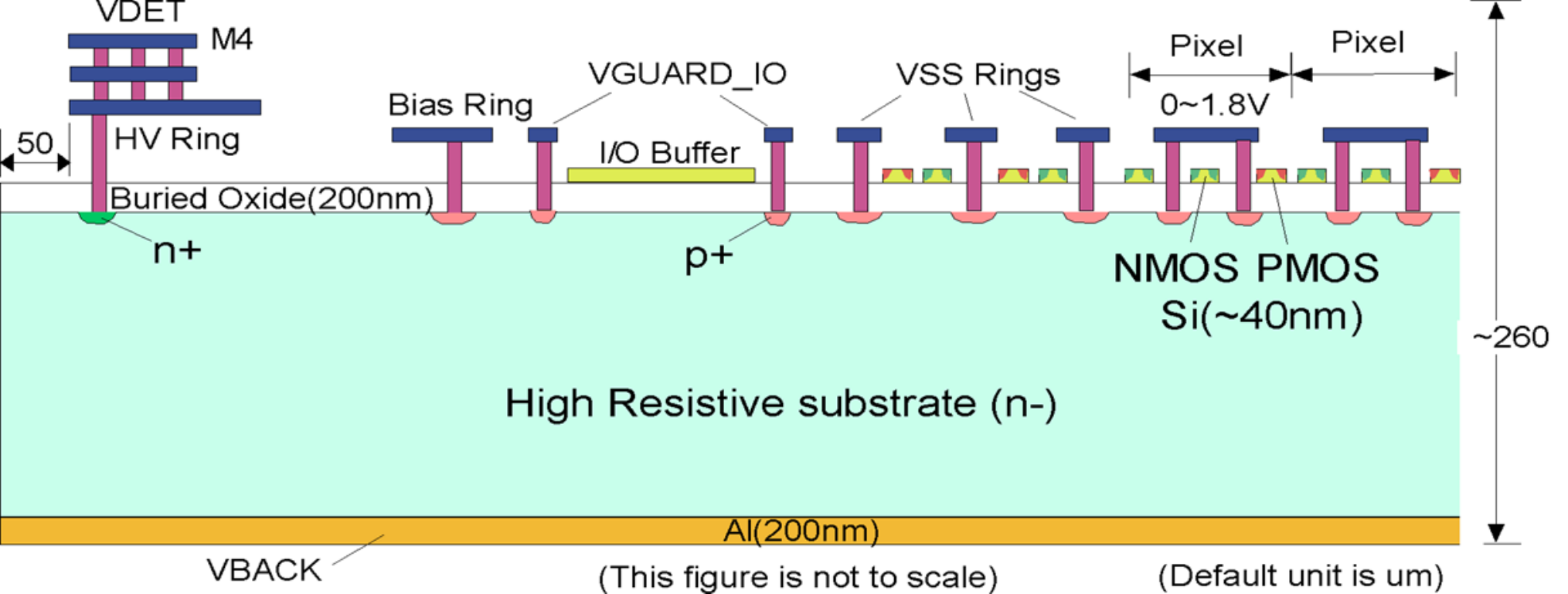}
    \label{fig:SOI1}}
    \vskip 0.3cm
\subfigure[SOI design variants ]{
    \includegraphics[width=0.4\textwidth]{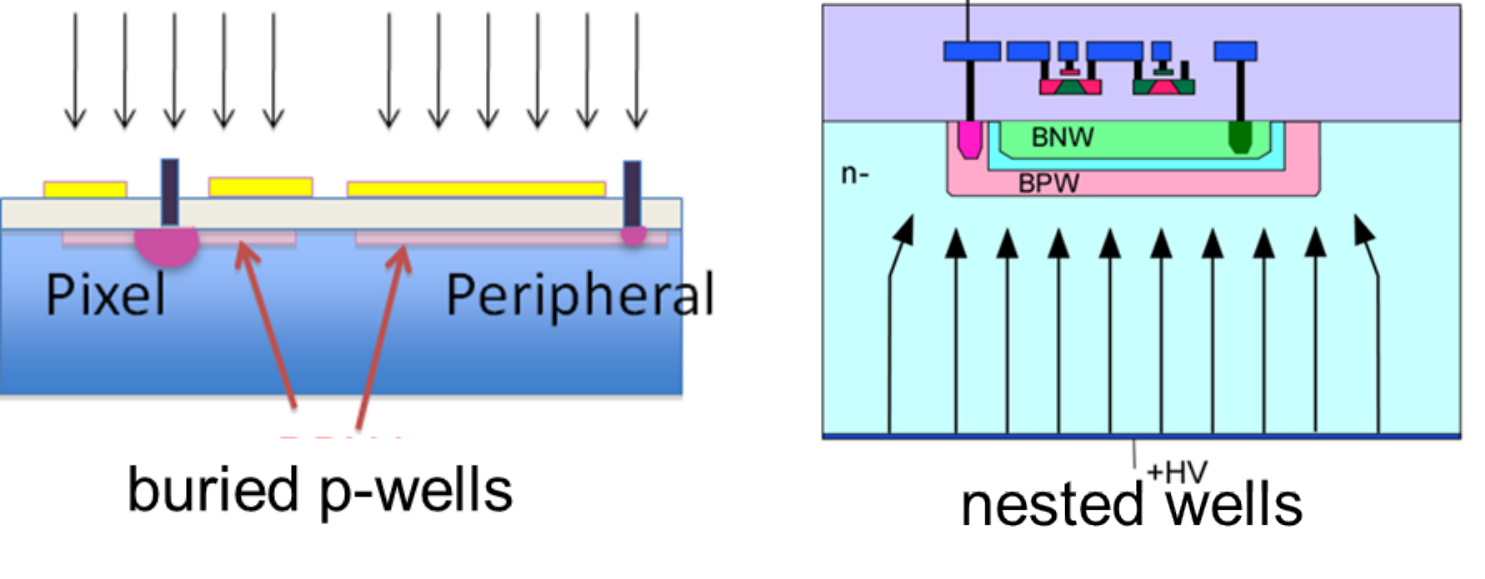}
    \label{fig:SOI2}}
    \vskip 0.3cm
 \subfigure[Double SOI structure]{
    \includegraphics[width=0.3\textwidth]{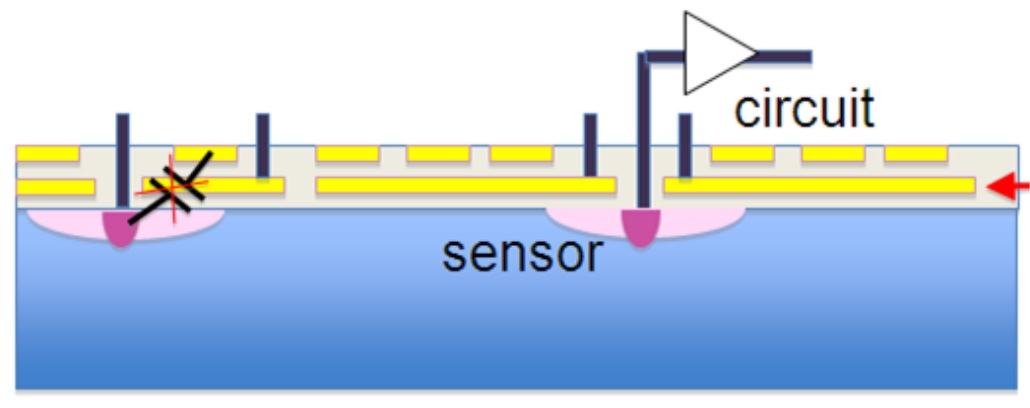}
    \label{fig:SOI3}}
\end{center}
\caption[]{\label{fig:SOI_variants} (a) Sketch of an SOI pixel detector: a CMOS layer is separated from the detection substrate by a the BOX; connection is made by vias. (b) Variants of the design introducing (left) buried p-wells, (right) nested wells, and (c) a double SOI structure, all designed to cope with problems (see text) experienced using the simple device~\cite{arai_pixel2010}.}
\end{figure}

The other important and promising approach to monolithic pixel devices is the Silicon-on-Insulator technology (SOI) progress in which was also reported at this conference~\cite{giubilato_pixel2010, arai_pixel2010, sato_pixel2010, hatsui_pixel2010}. The SOI technology promises full CMOS circuitry in the active area without
bump bonding with high sensitivity and full charge collection. Using an industrial process offers reliability and the smooth connection to the 3D integration technology which is currently of common interest in many groups. Figure~\ref{fig:SOI1} shows an example SOI
structure consisting of a high resistivity substrate, wafer bonded to a CMOS layer separated by a buried oxide layer (BOX) through which vias connect to the substrate. The main technical issues of the SOI technology are or have been (a) a reliable fabrication process, (b) the question of how to avoid the backgate effect (see below), (c) radiation hardness due to hole trapping inside the BOX, and (d) the
attainable resistivity of the substrate material. These are currently being addressed, especially at OKI~\cite{arai_pixel2010}.
As was shown at this conference, hole trapping in the box can to some extent be compensated by changing the substrate voltage~\cite{sato_pixel2010}. Several R$\&$D efforts have been reported~\cite{arai_pixel2010} done by OKI in collaboration with Fermilab to address in particular the backgate effect (by an additional buried p-well structure), cross talk (by implementing nested wells, i.e. deep buried p-well, buried n-well, or by a double SOI structure), and the radiation hardness (by double SOI layer wafers providing a biasable intermediate conducting layer that compensates built-up oxid charges), all sketched in Fig.~\ref{fig:SOI2} and \ref{fig:SOI3}. It is shown that indeed the backgate effect is suppressed by buried p-well structures. The next generation of tested devices and their performance in particle beams could be very interesting.
Finally, the experience of the past years with 3D integration efforts led by Fermilab was reported~\cite{yarema_pixel2010}. The various efforts with Tezzaron/Chartered, MIT-Lincoln Lab and OKI have shown that a long term commitment is needed for slow step-by-step successes. On the long run, however, a combination of monolithic devices
(SOI or CMOS active pixels) combined with 3D integration will probably be the path to go.

\section{Conclusions}
Pixel 2010 was the fifth very successful international conference in this series. It has presented the excellent performance of the LHC pixel detectors, the
progresses made in imaging applications, especially those at synchrotron light sources and XFELs, and has shown new interesting R$\&$D ideas and realizations that
will advance the pixel technology further in the coming years.
We are looking forward to the next conference, most likely in Japan.

\section*{Acknowledgments}
The author would like to thank the organizers of this wonderful conference with its excellent location and its spectacular surrounding, Andrei Starodumov, Danek Kotlinski and Roland Horisberger and all partici\-pants on whose contribution this summary paper is based.
%
\bibliographystyle{elsarticle-num}



\end{document}